\documentclass{svjour2} 
\smartqed 
\usepackage{graphicx}
\usepackage[T1]{fontenc}
\usepackage{amssymb, latexsym, amsopn}

\newcommand{\av}[1]{\langle{#1}\rangle_{\mathcal{D}}}

\begin{document} 

\title{Volume averaging in the quasispherical Szekeres model}

\author{Krzysztof Bolejko$^{1,2}$}

\institute{The University of Melbourne, Melbourne VIC 3010, Australia, and \\
Nicolaus Copernicus Astronomical Center,
	   Bartycka 18, 00-716 Warsaw, Poland \\
	  \email{bolejko@camk.edu.pl}}

\date{Received: date / Accepted: date}

\maketitle

\begin{abstract}
This paper considers  volume averaging in the quasispherical Szekeres model.
The volume averaging became of considerable interest after it was shown
that the volume acceleration calculated within the averaging framework
can be positive even when the local expansion rate decelerates.
This issue was intensively studied within spherically symmetric models.
However, since our Universe is not spherically symmetric similar analysis is needed 
in non-symmetrical models. This papers presents the averaging analysis
within the quasispherical Szekeres model which is a non-symmetrical generalisation of the
spherically symmetric Lema\^itre--Tolman family of models.
In the quasispherical Szekeres model the distribution of mass over
a surface of constant $t$ and $r$ has the form of a
mass-dipole superposed on a monopole. This paper shows that when calculating the volume  acceleration, $\ddot{a}$, within the Szekeres model, the dipole does not contribute to the final result, hence $\ddot{a}$ only depends on a monopole configuration. Thus, the volume averaging within the Szekeres model 
leads to literally the same solutions as those obtained within the Lema\^itre--Tolman model.
\end{abstract}

\PACS{98.80-k, 95.36.+x, 98.65.Dx}

\section{Introduction}
In the standard approach to cosmology it is assumed that the
Universe can be described by the homogeneous Friedmann model.
Within such a framework in order to correctly describe cosmological observations 
one needs to postulate the existence of dark energy, which in its simplest form
can be represented by a cosmological constant.
However, our Universe on scales up to at least 100 Mpc is very inhomogeneous.
Thus, it can evolve differently than the homogeneous model.
The difference between evolution of the homogeneous
model and the inhomogeneous Universe is known as a backreaction effect.
The direct study of the dynamical effects of inhomogeneities is difficult
since both general matter distribution and the numerical evolution of cosmological models
employing the full Einstein equations are unavailable at the level of
detail which would make them useful in studying this problem. 
Currently, therefore, one of the most popular approaches to backreaction is via averaging methods.

In the averaging approach to backreaction, one considers a solution to 
the Einstein equations for a general matter distribution and 
then an average of various observable quantities is taken. 
If several assumptions are introduced (see Sec. \ref{SBE}),
the averaging leads to the Buchert equations.
For a review on backreaction and the
Buchert averaging scheme the reader is referred to
\cite{R06,B08}.  
Within the Buchert averaging scheme 
there are a number of examples where it was shown by using 
spherically symmetric inhomogeneous models that one
can obtain negative values of the volume deceleration parameter
even if $\Lambda = 0$ \cite{NT05,PS06,KKNNY07,CGH08,S08,BA08}.

However, we should be aware that the results of averaging can be gauge-depended
\cite{KKM}. Another problem regarding an application
of spherical symmetric models is the problem with the age of the Universe.
Within the models studied in \cite{BA08} those of realistic density distribution and 
with $q<0$ had very large values of the $t_B$ function, of amplitude $10^{10}$ y (this means that the age of the Universe within such models is unrealistically small\footnote{This, however, does not apply 
to the two-scale averaging approach. For details on the two-scale models see works by R\"as\"anen \cite{SR06a}
or Wiltshire \cite{W07a,W07b,LNW08}.}).
This feature, however, can be an artefact of assumed spherical symmetry. 
Therefore, it is of great importance to study averaging in 
non-symmetrical inhomogeneous models. One of the immediate candidates is the Szekeres model \cite{S75a}.
The Szekeres model is a generalisation of the Lemaitre--Tolman model 
that has no symmetry \cite{BST77}.
Within the quasispherical Szekeres model one can describe two \cite{B06} or even three structures
\cite{B07}. Thus, the Szekeres model not only allows us to study how 
cosmic structures  affect their evolution
but also enables the examination of the volume acceleration of such systems which consist of several structures.
This paper, therefore, addresses the subject of volume averaging in the 
Szekeres model.

\section{Buchert equations}\label{SBE}

Since the Buchert averaging scheme involves the volume average
it applies to averaging of scalars only --
volume averaging of tensors leads to noncovariant quantities (for 
a review and a  detailed discussion about tensor averaging the reader is referred to \cite{AK97}). The average of a scalar $\Psi$ is then equal to 

\begin{equation}
\av{\Psi} = \frac{1}{V_{\mathcal{D}}} \int_{\mathcal{D}} {\rm d}^3x \sqrt{-h}
~ \Psi.
\end{equation}
where $h$ is a determinant of a 3D spatial metric, 
$h_{\alpha \beta} = g_{\alpha \beta} - u_{\alpha} u_{\beta}$;
${\mathcal{D}}$ is a domain of averaging, and $V_D$ its volume,

\begin{equation}
V_{\mathcal{D}}  =  \int_{\mathcal{D}} {\rm d}^3x \sqrt{-h}.
\end{equation}

The Buchert equations are obtained if the following are assumed

\begin{itemize}
\item  the Universe is filled with an irrotational dust only,
\item the metric is of the form $ds^2 = dt^2 - g_{ij} dx^i dx^j$
(3+1 ADM space-time foliation with a constant lapse and a 
vanishing shift vector).
\end{itemize}
Then by averaging the Raychaudhuri equation we obtain \cite{B00}

\begin{equation}
3 \frac{\ddot{a}_{\mathcal{D}}}{a_{\mathcal{D}}} = - 4 \pi \av{\rho} + \mathcal{Q}_{\mathcal{D}},
\label{bucherteq1}
\end{equation}
where a dot ($\dot{}$) denotes $\partial_t$; the scale factor $a_{\mathcal{D}}$ is defined as a cube root of the volume

\begin{equation}
a_{\mathcal{D}} = (V_{\mathcal{D}}/V_{\mathcal{D}_i})^{1/3},
\label{aave}
\end{equation}
(where $V_{\mathcal{D}_i}$ is an initial volume); and the 
backreaction term $\mathcal{Q}_{\mathcal{D}}$ is given by
\begin{equation}
\mathcal{Q}_{\mathcal{D}} \equiv \frac{2}{3}\left( \av{{\Theta^2}} - \av{ \Theta }^2 \right)
- 2 \av{ \sigma^2},
\label{qdef}
\end{equation}
where $\Theta$ is the scalar of expansion and $\sigma$ is the shear
scalar.
Averaging of the Hamiltonian constraint leads to \cite{B00}

 \begin{equation}
3 \frac{\dot{a}_{\mathcal{D}}^2}{a_{\mathcal{D}}^2} = 8 \pi  \av{ \rho}	-
\frac{1}{2} \av{ \mathcal{R} } - \frac{1}{2} \mathcal{Q}_{\mathcal{D}}, \label{bucherteq2} 
\end{equation}
where $\av{ \mathcal{R} }$ is an average of the spatial Ricci scalar $^{(3)}
\mathcal{R}$. The above is compatible with (\ref{bucherteq1}) if the
integrability condition holds

\begin{equation}
\frac{1}{a_{\mathcal{D}}^6} \partial_t \left( \mathcal{Q}_{\mathcal{D}} a_{\mathcal{D}}^6 \right) + 
\frac{1}{a_{\mathcal{D}}^2} \partial_t \left( \av{R} a_{\mathcal{D}}^2 \right)  = 0.
\label{intcond} 
\end{equation}

Equations (\ref{bucherteq1}) and (\ref{bucherteq2}) are very similar to the
Friedmann equations, where $\mathcal{Q}_{\mathcal{D}}=0$, and $\rho$ and $\mathcal{R}$ depend on time only.  
In fact, it can be shown that they are kinematically equivalent to  Friedmann equations with a scalar field \cite{BLA06}. 

Using (\ref{bucherteq1}) and (\ref{bucherteq2}) we can calculate the deceleration parameter 

\begin{equation}
q \equiv - \frac{\ddot{a}_{\mathcal{D}}a_{\mathcal{D}}}{\dot{a}_{\mathcal{D}}^2}  = - \frac{- 4 \pi G \av{ \rho } + \mathcal{Q}_{\mathcal{D}}}{8 \pi G \av{ \rho }  -
\frac{1}{2} \av{ \mathcal{R} } - \frac{1}{2} \mathcal{Q}_{\mathcal{D}}}.
\label{decparave}
\end{equation}
In the standard approach to cosmology where Friedmann models are employed the case
of $q<0$ implies that $\Lambda > 0$. However, as it was shown
within inhomogeneous but isotropic models \cite{NT05,PS06,KKNNY07,CGH08,S08,BA08}
 we can have $q<0$ even if  $\Lambda = 0$.
This suggest that the apparent acceleration of the Universe might be
explained not by invoking dark energy but by taking into account
matter inhomogeneities.
In Sec. \ref{sec4} we will examine this issue by employing
the non-symmetrical Szekeres model.

\section{The Szekeres model}\label{SZsec}

The metric of the Szekeres model is of the following form  \cite{S75a}

\begin{equation}
{\rm d} s^2 = {\rm d} t^2 - X^2 {\rm d} r^2 - Y^2 ( {\rm d}x^2 + {\rm d}y^2). \label{ds2}
\end{equation}
For our purpose it is more  convenient to adopt a pair of complex conjugate coordinates

\begin{equation}
\zeta = x + i y, \quad \bar{\zeta} = x - i y,
\end{equation}
so that the metric becomes

\begin{equation}
{\rm d} s^2 = {\rm d} t^2 - X^2 {\rm d} r^2 - Y^2 {\rm d} \zeta {\rm d} \bar{\zeta}. \label{ds21}
\end{equation}
where
\[ X = \frac{{\cal E}(r,\zeta,\bar{\zeta}) Y'(t,r,\zeta,\bar{\zeta})}{\sqrt{\varepsilon-k(r)}}, \quad Y = \frac{\Phi(t,r)}{{\cal E}(r,\zeta,\bar{\zeta})}, \]
and
\[ {\cal E}= a(r) \zeta \bar{\zeta} + b(r) \zeta + c(r) \bar{\zeta} + d(r), \quad \varepsilon = 0, \pm 1. \]
Here a prime (') denotes $\partial_r$.

The case where $\varepsilon = -1$ is often called the  
quasihyperbolic Szekeres model (for a detailed discussion on the 
quasihyperbolic Szekeres models see \cite{HK08}),
$\varepsilon = 0$ quasiplane  (for details see \cite{HK08,K08}),
and $\varepsilon = 1$ quasispherical (for details see \cite{HK02}).
Although it is possible to have within one model quasispherical  and 
quasihyperbolic regions separated by the quasiplane region \cite{HK08},
only the quasispherical case will be considered here.
This is because the averaging within the 
quasihyperbolic and quasiplane requires a special treatment.
Firstly, an area of a surface of constant $t$ and $r$ 
in the quasihyperbolic and quasiplane models is infinite. Secondly, there is no origin --
in the quasihyperbolic model $r$ cannot be equal to $0$,
and in the quasiplane $r$ can only asymptotically approach 
the origin, $r \rightarrow 0$ \cite{HK08}.

In the quasispherical Szekeres model a surface of constant $t$ and $r$
is a sphere of radius $\Phi(r,t)$  \cite{S75a}. Thus, the quasispherical Szekeres model 
is a generalisation of the Lema\^itre--Tolman model \cite{L33,T34}.
Within the Szekeres model shells of matter, however, are not concentric.
The quasispherical Szekeres model becomes the Lema\^itre--Tolman (LT) model when ${\cal E}'=0$.
In this case

\[ X = \frac{\Phi'}{\sqrt{1-k(r)}}, \quad Y = \frac{\Phi}{{\cal E}}, \quad \frac{{\rm d} \zeta {\rm d} \bar{\zeta}}{{\cal E}^2} = {\rm d}{\theta}^2 + \sin^2\theta {\rm d} {\phi}^2. \]

Originally, Szekeres considered only a case of $p=0=\Lambda$.
This result was generalised by Szafron \cite{Szf77} to the case of uniform pressure, $p=p(t)$. 
A special case of this solution, the cosmological constant, was discussed 
in detail by Barrow and Stein-Schabes \cite{BSS84}.
In the case of $p=0=\Lambda$, the Einstein equations reduce to

\begin{equation}
\dot{\Phi}^2  = \frac{2M}{\Phi} - k,
\label{vel}
\end{equation}

\begin{equation}
4 \pi \rho = \frac{M' - 3 M {\cal E}'/{\cal E}}{\Phi^2 ( \Phi' - \Phi {\cal E}'/{\cal E})},
\label{rho}
\end{equation}
where $M(r)$ is another arbitrary function. 
In a Newtonian limit $M$ is equal to the mass inside the shell of 
radial coordinate $r$.  Although the $\rho$ function, as seen from  (\ref{rho}),  is a function of all coordinates,
it can be shown that the distribution of mass over each
single sphere of constant $t$ and $r$
has the structure of a mass dipole superposed on a monopole,
$\rho(t,r,\zeta, \bar{\zeta}) = \rho_{mon}(t,r) + \rho_{dip}(t,r,\zeta, \bar{\zeta})$ 
\cite{S75b,dS85,PK06}. In general case, the orientation
of the dipole axis is different on every constant-($t,r$) sphere.

The dipole contribution vanishes when ${\cal E}'=0$ and then the Szekeres model
reduces to the LT model.

The scalar of expansion is equal to
\begin{equation}
\Theta =  u^{\alpha}{}_{;\alpha} = 
\frac{\dot{\Phi}' + 2 \dot{\Phi} \Phi'/ \Phi
- 3 \dot{\Phi} {\cal E}'/{\cal E}}{
\Phi' - \Phi {\cal E}'/{\cal E}}.
\end{equation}
The scalar of shear is

\begin{equation}
\sigma^2 = \frac{1}{3} \left( \frac{\dot{\Phi}' - \dot{\Phi} \Phi'/ \Phi}{\Phi' - \Phi {\cal E}'/{\cal E}} \right)^2.
\end{equation}
The spatial Ricci scalar $^{(3)}\mathcal{R}$ is equal to

\begin{equation}
^{(3)}\mathcal{R} =
2 \frac{k}{\Phi^2} \left( \frac{ \Phi k'/k - 2 \Phi  {\cal E}'/{\cal E}}{
\Phi' - \Phi {\cal E}'/{\cal E}} + 1 \right).
\label{qsz3R}
\end{equation}
In the LT limit these scalars reduce to

\begin{equation}
\Theta =  \frac{\dot{\Phi}'}{\Phi'} + 2 \frac{\dot{\Phi}}{\Phi},
\quad
\sigma^2 = \frac{1}{3} \left( \frac{\dot{\Phi}'}{\Phi'} - \frac{\dot{\Phi}}{\Phi} \right)^2,
\quad
^{(3)}\mathcal{R} =  2 \frac{( \Phi k )'}{\Phi^2 \Phi'}.
\label{thtlt}
\end{equation}

\section{Averaging in the quasispherical Szekeres model}\label{sec4}

This section considers
the volume averaging within the quasispherical Szekeres model.
The volume is calculated around the observer located at the origin.
It will be shown that the dipole configuration does not affect
such quantities as $\av{\rho}$, $\mathcal{Q}_{\mathcal{D}}$, or $\av{\mathcal{R}}$.
These functions only depend on the monopole configuration.
First let us notice that the volume in the Szekeres model is exactly  the same as in the
LT model (i.e. if ${\cal E}' = 0$). Following the method presented in \cite{S75b}
we obtain

\begin{eqnarray}
&& V_{\mathcal{D}} =  \int\limits\limits_0^{r_{\mathcal{D}}} {\rm d} {r} \int \int {\rm d} \zeta {\rm d} \bar{\zeta} ~
X Y^2  = \int\limits_0^{r_{\mathcal{D}}} {\rm d} {r} \int \int {\rm d} \zeta {\rm d} \bar{\zeta}
\frac{\Phi^2}{\sqrt{1-k}} \left( \Phi' - \Phi \frac{{\cal E}'}{{\cal E}} \right) \frac{1}{{\cal E}^2}  \nonumber \\
&& = 
\int\limits_0^{r_{\mathcal{D}}} {\rm d} {r}
\left[ \frac{\Phi^2 \Phi'}{\sqrt{1-k}} 
\int \int \frac{{\rm d} \zeta {\rm d} \bar{\zeta}}{{\cal E}^2} +
\frac{1}{2} \frac{\Phi^3 }{\sqrt{1-k}} \frac{\partial}{\partial  {r}} \left( 
 \int \int \frac{{\rm d} \zeta {\rm d} \bar{\zeta}}{{\cal E}^2} \right) \right].
\end{eqnarray}

\noindent Since ${\rm d} \zeta {\rm d} \bar{\zeta}/{\cal E}^2$ is the metric of a unit sphere,
hence

\[ \int \int \frac{{\rm d} \zeta {\rm d} \bar{\zeta}}{{\cal E}^2} = 4 \pi. \]

\noindent Thus 

\begin{equation}
V_{\mathcal{D}} = 4 \pi \int\limits_0^{r_{\mathcal{D}}} {\rm d}  {r} \frac{\Phi^2 \Phi'}{\sqrt{1-k}}. 
\label{vol}
\end{equation}
The same result is obtained if initially ${\cal E}'$ is set to zero.
Thus, the dipole component does not contribute to total volume.
As it will be shown below it also does not contribute to $\av{\rho}$.

\begin{eqnarray}
&& \av{\rho} = \frac{1}{V_{\mathcal{D}}} \int\limits_0^{r_{\mathcal{D}}} {\rm d}  {r} \int \int {\rm d} \zeta {\rm d} \bar{\zeta} ~
X Y^2 \rho  \nonumber \\
&& = \frac{1}{4 \pi V_{\mathcal{D}}} \int\limits_0^{r_{\mathcal{D}}} {\rm d}  {r} \int \int \frac{{\rm d} \zeta {\rm d} \bar{\zeta}}{{\cal E}^2} \frac{\Phi^2}{\sqrt{1-k}}
\left( \Phi' - \Phi \frac{{\cal E}'}{{\cal E}} \right)
\frac{M' - 3 M {\cal E}'/{\cal E}}{\Phi^2 \left( \Phi' - \Phi {\cal E}'/{\cal E} \right)}  \nonumber \\
&& = \frac{1}{V_{\mathcal{D}}} \int\limits_0^{r_{\mathcal{D}}} {\rm d}  {r} \frac{M'}{\sqrt{1-k}}  + 
\frac{3}{8 \pi V_{\mathcal{D}}} \int\limits_0^{r_{\mathcal{D}}} {\rm d}  {r} \frac{M}{\sqrt{1-k}}
 \frac{\partial}{\partial  {r}} \left( 
 \int \int \frac{{\rm d} \zeta {\rm d} \bar{\zeta}}{{\cal E}^2} \right)  \nonumber \\
&& = \frac{1}{V_{\mathcal{D}}} \int\limits_0^{r_{\mathcal{D}}}{\rm d}  {r} \frac{M'}{\sqrt{1-k}}.
\label{arho}
\end{eqnarray}

\noindent The average of the scalar of the expansion is

\begin{eqnarray}
&& \av{\Theta} =  \frac{1}{V_{\mathcal{D}}} \int\limits_0^{r_{\mathcal{D}}}{\rm d}  {r} \int \int \frac{ {\rm d} \zeta {\rm d} \bar{\zeta}}{{\cal E}^2}
\frac{\Phi^2}{\sqrt{1-k}} \left( \Phi' - \Phi \frac{{\cal E}'}{{\cal E}} \right)  \frac{\dot{\Phi'} + 2 \dot{\Phi}{\Phi'}/ \Phi - 3 \dot{\Phi} {\cal E}'/{\cal E}}{\Phi' - \Phi {\cal E}'/{\cal E}}   \nonumber \\ 
&& = \frac{4 \pi}{V_{\mathcal{D}}} \int\limits_0^{r_{\mathcal{D}}}{\rm d}  {r} 
 \frac{\Phi^2 \Phi'}{\sqrt{1-k}} \left(
 \frac{\dot{\Phi}'}{\Phi'} + 2 \frac{\dot{\Phi}}{\Phi} \right).
\label{atht}
\end{eqnarray}
As above, ${\cal E}'/{\cal E}$  does not contribute to the final result and,
as seen from (\ref{thtlt}), the result is the same as in the LT model.
To calculate the backreaction term $\mathcal{Q}_{\mathcal{D}}$ we still need to 
find the average of the following quantity

\begin{eqnarray}
&& \frac{2}{3} \av{\Theta^2} - 2 \av{\sigma^2} = \frac{2}{3} \frac{1}{V_{\mathcal{D}}} \int\limits_0^{r_{\mathcal{D}}}{\rm d}  {r} \int \int \frac{ {\rm d} \zeta {\rm d} \bar{\zeta}}{{\cal E}^2} 
\frac{\Phi^2}{\sqrt{1-k}} \left( \Phi' - \Phi \frac{{\cal E}'}{{\cal E}} \right) \nonumber \\
&& \times \frac{ \left( \dot{\Phi}' + 2 \dot{\Phi} \Phi' / \Phi -  3 \dot{\Phi} {\cal E}'/{\cal E}
\right)^2 - \left( \dot{\Phi}' - \dot{\Phi} \Phi' / \Phi \right)^2}{
\left( \Phi' - \Phi {\cal E}'/{\cal E} \right)^2}
=  \nonumber \\
&& = \frac{2}{3} \frac{1}{V_{\mathcal{D}}} \int\limits_0^{r_{\mathcal{D}}}{\rm d}  {r}
\int \int \frac{ {\rm d} \zeta {\rm d} \bar{\zeta}}{{\cal E}^2} 
\frac{\Phi^2}{\sqrt{1-k}}
\left( 3 \frac{\dot{\Phi}^2}{\Phi^2} \Phi' + 6 \frac{\dot{\Phi} \dot{\Phi}'}{\Phi} 
+ 3 \frac{\dot{\Phi}^2}{\Phi^2} \Phi \frac{{\cal E}'}{{\cal E}} - 12 \frac{\dot{\Phi}^2}{\Phi} \frac{{\cal E}'}{{\cal E}} \right) \nonumber \\
&& = \frac{8 \pi}{3} \frac{1}{V_{\mathcal{D}}} \int\limits_0^{r_{\mathcal{D}}}{\rm d}  {r}
\frac{\Phi^2 \Phi'}{\sqrt{1-k}} 
\left( 3 \frac{\dot{\Phi}^2}{\Phi^2}  + 6 \frac{\dot{\Phi} \dot{\Phi}'}{\Phi \Phi'}
\right).
\label{tms2} 
\end{eqnarray}

\noindent This result is also the same if initially  ${\cal E}'$ is set to zero.
Finally let us notice that $\av{\mathcal{R}}$ is also the same as in the case of ${\cal E}'=0$.
Averaging relation (\ref{qsz3R}) yields

\begin{eqnarray}
&& \av{\mathcal{R}} =  \frac{1}{V_{\mathcal{D}}} \int\limits_0^{r_{\mathcal{D}}}{\rm d}  {r} \int \int \frac{ {\rm d} \zeta {\rm d} \bar{\zeta}}{{\cal E}^2}
 \left( \Phi' - \Phi \frac{{\cal E}'}{{\cal E}} \right)  \frac{2 k}{\sqrt{1-k}} 
  \left( \frac{ \Phi k'/k - 2 \Phi  {\cal E}'/{\cal E}}{
\Phi' - \Phi {\cal E}'/{\cal E}} + 1 \right)   \nonumber \\ 
 && =  \frac{8 \pi}{V_{\mathcal{D}}} \int\limits_0^{r_{\mathcal{D}}}{\rm d}  {r} \frac{( \Phi k )'}{\sqrt{1-k}} .
\label{a3r}
\end{eqnarray}

Inserting (\ref{arho})--(\ref{a3r}) into (\ref{bucherteq1}) and (\ref{bucherteq2})
we see that the dipole configuration contributes neither to $\dot{a}_{\mathcal{D}}$ nor to $\ddot{a}_{\mathcal{D}}$. The volume acceleration, 
as well as the volume deceleration parameter, depends
only on the monopole distribution ---
it depends only on functions of $t$ and $r$ [i.e. $M(r)$, $k(r)$ and $\Phi(t,r)$],
and not on $\zeta$ and $\bar{\zeta}$ [the dependence on
${\cal E}(t,r,\bar{\zeta},{\zeta})$) vanishes].
Moreover,  $\av{\rho}$, $\mathcal{Q}_{\mathcal{D}}$, and $\av{\mathcal{R}}$ are 
literally of the same form as if initially  ${\cal E}'$ is set to zero (the LT case).

\section{Conclusions}

In this paper the volume averaging within the quasispherical Szekeres model has been investigated. The Szekeres model is a generalisation of the LT model. 
The density distribution in the quasispherical Szekeres model has the structure of a time-dependent mass dipole superposed on a monopole.
When calculating the volume  acceleration ($\ddot{a}$ to be more exact) 
or volume deceleration parameter ($q$)
within the quasispherical Szekeres, 
the dipole does not contribute to the final result and  $\ddot{a}$ only depends on a monopole configuration.
The solutions are the same if initially ${\cal E}'$ was set to zero, thus the 
results and conclusions found  when studying averaging
within the LT models also apply to the Szekeres models.
For example, we can, without any further calculations, conclude
that $\mathcal{Q}_{\mathcal{D}} =0$ when $k=0$. This is an implication
of the result obtained by Paranjape and Singh \cite{PS06} who showed that in the parabolic ($k=0$) LT model
the backreaction term, $\mathcal{Q}_{\mathcal{D}}$,  vanishes.
Another result obtained within the LT models which, 
as has been shown, also appears to apply to the Szekeres models
is that $\mathcal{Q}_{\mathcal{D}} >0$ is only possible for unbounded systems, $k<0$ \cite{S08}.
However, as shown in \cite{BA08}, in most cases this also requires
that the bang time function, $t_B$, is of  amplitude of $10^{10}$ y.

\begin{acknowledgements}
This research has been supported by The Peter and Patricia Gruber Foundation and the International Astronomical Union.
I thank Paulina Wojciechowska, Catherine Buchanan 
and Henk van Elst and the referees for their useful comments and suggestions.

\end{acknowledgements}

\end{document}